\documentclass[aps,preprint,prd,nopacs,nofootinbib]{revtex4}

\usepackage{amsmath}
\usepackage{graphicx}
\usepackage{siunitx}
\usepackage{mathtools}
\usepackage{makecell}
\usepackage[colorlinks,linkcolor=magenta,anchorcolor=cyan,citecolor=blue]{hyperref}

\begin{document}
\title{Testing the $n_s-H_0$ scaling relation with Planck-independent CMB data}

\author{Ze-Yu Peng$^{1,2}$\footnote{pengzeyu23@mails.ucas.ac.cn}}
\author{Yun-Song Piao$^{1,2,3,4}$\footnote{yspiao@ucas.ac.cn}}

\affiliation{$^1$ School of Physical Sciences, University of
Chinese Academy of Sciences, Beijing 100049, China}

\affiliation{$^2$ International Centre for Theoretical Physics
Asia-Pacific, University of Chinese Academy of Sciences, 100190
Beijing, China}

\affiliation{$^3$ School of Fundamental Physics and Mathematical
    Sciences, Hangzhou Institute for Advanced Study, UCAS, Hangzhou
    310024, China}

\affiliation{$^4$ Institute of Theoretical Physics, Chinese
    Academy of Sciences, P.O. Box 2735, Beijing 100190, China}

\begin{abstract}

In early dark energy (EDE) resolution of Hubble tension, the spectral index $n_s$ of primordial scalar perturbation follows a scaling relation ${\delta n_s}\simeq 0.4\frac{\delta H_0}{H_0}$, where $H_0$ is the Hubble constant. However, this $n_s-H_0$ relation was obtained based on the datasets including Planck cosmic microwave background (CMB) data. In this paper, we investigate this scaling relation with Planck-independent CMB data, i.e. ACT and SPT-3G combined with WMAP(+BAO+Pantheon), respectively. Our results show that the WMAP+SPT-3G dataset also follows this scaling relation, while the WMAP+ACT dataset seems to favor smaller $n_s$, which is related to the fact that the critical redshift $z_c$, at which EDE is excited, favored by the WMAP+ACT dataset is lower and closer to the recombination time. 

\end{abstract}

\maketitle


\section{Introduction}

The spectral index $n_s$ of primordial scalar perturbation  plays a crucial role in understanding the physics of inflation. It is well-known that the combination of Planck cosmic microwave background (CMB) data with other datasets favored $n_s = 0.965\pm 0.004$ (68\% CL) \cite{Planck:2018vyg}, which, however, is based on the $\Lambda$CDM model.

In recent years, with the accumulation of observational data, the standard cosmological model, $\Lambda$CDM, is confronted with increasingly severe challenges, among which the most significant is the so-called Hubble tension
(see e.g. \cite{Abdalla:2022yfr,DiValentino:2021izs} for recent reviews). It widely exists among different measurements, which seems unlikely to be entirely resolved by systematic errors, so currently it has arrived at a consensus that new physics beyond $\Lambda$CDM might be required \cite{Mortsell:2018mfj,DiValentino:2021izs,Vagnozzi:2019ezj,Knox:2019rjx}.

In the pre-combination resolutions of the Hubble tension, such as early dark energy (EDE) \cite{Karwal:2016vyq,Poulin:2018cxd}, an unknown EDE component in the energy density of the Universe, which is non-negligible only for a short epoch before recombination, suppresses the sound horizon $r_s=\int \frac{c_s}{H(z)}dz$ before recombination, where $c_s$ is the sound speed, $H$ is the Hubble parameter, and $z$ denotes redshift.
As a result, the CMB observations will prefer a higher Hubble constant $H_0$. Recently, a variety of phenomenological EDE models has been proposed in e.g. \cite{Agrawal:2019lmo,Lin:2019qug,Niedermann:2019olb,Sakstein:2019fmf,Kaloper:2019lpl,Alexander:2019rsc,Berghaus:2019cls,Ye:2020btb,Ye:2020oix,Ye:2021iwa,Braglia:2020bym,Seto:2021xua,Karwal:2021vpk,Clark:2021hlo,Nojiri:2021dze,Nojiri:2022ski,MohseniSadjadi:2022pfz,Sabla:2022xzj,Wang:2022jpo,Wang:2022nap,Reeves:2022aoi,Gomez-Valent:2022bku,Ye:2023zel,Brissenden:2023yko,Nojiri:2023mvi,Odintsov:2023cli,Liu:2023kce,Ben-Dayan:2023rgt}, see also Ref.~\cite{Poulin:2023lkg} for a comprehensive review.

In Ref.~\cite{Ye:2021nej}, it has been found that in such pre-recombination models, the fullPlanck+BAO+Pantheon+R19 dataset imposes the shift of $n_s$ scale as
\begin{equation} \label{eq:deltans}
    {\delta n_s}\simeq 0.4\frac{\delta H_0}{H_0},
\end{equation}
which thus suggests a scale-invariant Harrison-Zeldovich spectrum of primordial scalar perturbation, i.e. $n_s=1$ ($|n_s-1|\sim {\cal O}(0.001)$) for $H_0\sim73$km/s/Mpc, see also \cite{Jiang:2022uyg,Jiang:2022qlj}. This result has also been exhibited clearly earlier in AdS-EDE model \cite{Ye:2020btb}, see also \cite{DiValentino:2018zjj,Giare:2022rvg,Calderon:2023obf,Giare:2023wzl} for studies on the correlation between $n_s=1$ and possible resolutions of the Hubble tension. The constraint on tenor-to-scalar ratio $r$ with up-to-date BICEP/Keck data has been recently considered in Ref.~\cite{Ye:2022afu,Jiang:2023bsz,Cruz:2022oqk}.


However, it's well known that the Planck data suffers from some minor issues, such as the so-called $A_{\mathrm{lens}}$ anomaly \cite{Calabrese:2008rt} and the inconsistency between high-$\ell$ and low-$\ell$ Planck's TT power spectra \cite{Addison:2015wyg,Planck:2016tof}. Therefore, it is significant to test the $n_s-H_0$ relation (\ref{eq:deltans}) with CMB data other than Planck, especially ground-based datasets like ACT \cite{ACT:2020frw,ACT:2020gnv} and SPT \cite{SPT-3G:2021eoc,SPT-3G:2022hvq}, which are more precise on small scales than Planck.

Recently, it has been found that the fourth data release of ACT \cite{ACT:2020frw,ACT:2020gnv}, combined with WMAP\cite{WMAP:2012fli} or restricted Planck ($\ell_{TT}\lesssim 1000$) data, signals a support for EDE at $2-3\sigma$, without including any $H_0$ prior \cite{Hill:2021yec,Poulin:2021bjr,Jiang:2022uyg,Simon:2022adh}, see also \cite{LaPosta:2021pgm,Jiang:2021bab,Smith:2022hwi,Jiang:2022uyg} for the inclusion of SPT or SPT-3G data \cite{SPT-3G:2021eoc}. What's more, an $n_s-H_0$ relation similar to (\ref{eq:deltans}) has also been shown with Planck($\ell_{TT}\lesssim 1000$)+ACT+SPT+BAO+Pantheon \cite{Jiang:2022uyg,Smith:2022hwi}.

In this paper, we will test the $n_s-H_0$ scaling relation (\ref{eq:deltans}) with Planck-independent CMB data, i.e. ACT and SPT-3G combined respectively with WMAP. The details of our MCMC analysis are provided in section-\ref{sec:methods}. Our analysis results are shown in section-\ref{sec:results}, and discussed in section-\ref{sec:discussion}. We conclude in section-\ref{sec:conclusions}. Additionally, we briefly describe the EDE models we consider in appendix-\ref{app:models}, and show the MCMC results of $\Lambda$CDM in appendix-\ref{app:LCDM} as a comparison.

\section{Data sets and methods}\label{sec:methods}

The CMB datasets we consider are:
\begin{itemize}
    \item \textbf{WMAP}: The python adapted likelihood of WMAP\footnote{\url{https://github.com/HTJense/pyWMAP}}. This likelihood corresponds to the WMAP 9-year CMB observations, including TT power spectrum with multipole range $2\le\ell\le 1200$ and TE power spectrum with multipole range $24\le\ell\le 800$ \cite{WMAP:2012fli}. Here, following Ref.~\cite{ACT:2020gnv}, we do not use the low-$\ell$ ($2\le\ell\le 23$) polarization likelihood, which is replaced by a Gaussian prior on the optical depth, $\tau_{\mathrm{reio}} = 0.065\pm0.015$.
    \item \textbf{SPT-3G}: The recently updated SPT-3G 2018 TT/TE/EE likelihood\footnote{\url{https://github.com/xgarrido/spt_likelihoods}}. This likelihood corresponds to the CMB data obtained from the observations of an approximately $1500\mathrm{~deg^2}$ region in the southern sky made by SPT-3G in 2018, including TE and EE power spectra with multipole range $300<\ell\le 3000$ and TT power spectrum with mutipole range $750<\ell\le 3000$ \cite{SPT-3G:2022hvq}. \item \textbf{ACT}: The ACTPol DR4 CMB Power Spectrum Likelihood\footnote{\url{https://github.com/ACTCollaboration/pyactlike}}. This likelihood corresponds to the CMB data measured by the Atacama Cosmology Telescope from $5400~\mathrm{deg^2}$ of the 2013–2016 survey, including TE and EE power spectra with multipole range $350.5\le\ell\le 4125.5$ and TT power spectrum with mutipole range $600.5\le\ell\le 4125.5$ \cite{ACT:2020frw,ACT:2020gnv}.
\end{itemize}
In this work, we will separately combine the ACT and SPT-3G datasets with WMAP. And in our two combined datasets, we will also include the BAO measurements from 6dFGS at $z=0.106$ \cite{beutler20116df}, SDSS DR7 at $z = 0.15$ \cite{Ross:2014qpa}, and BOSS DR12 at $z = 0.38, 0.51, 0.61$ \cite{BOSS:2016wmc}, as well as the uncalibrated luminosity distance of Type Ia supernovae from the Pantheon sample, with redshift ranging $0.01<z<2.3$ \cite{Pan-STARRS1:2017jku}.

The injection of EDE before recombination can raise $n_s$, so we will confront the axion-like EDE and AdS-EDE models with our datasets to test the $n_s-H_0$ relation, as in Refs.\cite{Ye:2021nej,Jiang:2022uyg}. Here, for the six standard $\Lambda$CDM parameters, we take uninformative flat priors on parameters $\left\{\ln(10^{10}A_s),n_s,H_0,\omega_b,\omega_{\mathrm{cdm}}\right\}$ and a Gaussian prior on the optical depth, $\tau_{\mathrm{reio}}=0.065\pm0.015$. Also, we set the prior of axion-like EDE parameters as: $f_{\mathrm{EDE}}\in [0,0.3]$, $\log_{10}{z_c}\in [3,4]$, $\theta_i\in[0,3.1]$, and the prior of AdS-EDE parameters as: $f_{\mathrm{EDE}}\in [0,0.3]$, $\log_{10}{z_c}\in [3,4]$, see Appendix-\ref{app:models} for both models and relevant parameters. In AdS-EDE, we fix $\alpha_{\mathrm{AdS}} \equiv\left(\rho_{\mathrm{m}}\left(z_c\right)+\rho_{\mathrm{r}}\left(z_c\right)\right) V_{\mathrm{AdS}}=3.79 \times 10^{-4}$ as in Ref.~\cite{Ye:2020btb}.

To perform MCMC sampling, we use the publicly available code \texttt{Cobaya} \cite{Torrado:2020dgo}\footnote{\url{https://github.com/CobayaSampler/cobaya}}, with an adaptive, speed-hierarchy-aware MCMC sampler \cite{Lewis:2002ah,Lewis:2013hha} and the fast-dragging procedure described in \cite{Neal:2005}. The models are calculated using \texttt{CLASS} \cite{Blas:2011rf} and its modified version\footnote{The codes are available at \url{https://github.com/PoulinV/AxiCLASS} for axion-like EDE and \url{https://github.com/genye00/class_multiscf} for AdS-EDE.}. We take our MCMC chains to be converged using the Gelman-Rubin criterion \cite{10.1214/ss/1177011136} with $R-1<0.1$. The posterior distribution is plotted using \texttt{GetDist} \cite{Lewis:2019xzd}. The best-fit parameters are obtained using the \texttt{Py-BOBYQA} implementation \cite{2018arXiv180400154C,2018arXiv181211343C} of the \texttt{BOBYQA} minimization algorithm \cite{BOBYQA}.

\section{Results\label{sec:results}}
\subsection{axion-like EDE}
\begin{table}
    \centering
    \begin{tabular}{|c|c|c|c|}
    \hline
    Parameters & \makecell[c]{WMAP+ACT\\+BAO+Pantheon} & \makecell[c]{WMAP+SPT-3G\\+BAO+Pantheon} & \makecell[c]{Planck\\+BAO+Pantheon} \\
    \hline\hline
    $f_{\mathrm{EDE}}$ & $0.133(0.245)_{-0.082}^{+0.039}$ & $<0.119(0.021)$ & $<0.091(0.088)$ \\
    $\log_{10}(z_c)$ & $3.30(3.48)_{-0.13}^{+0.20}$ & unconstrained $(3.977)$ & unconstrained $(3.55)$ \\
    $\Theta_i$ & unconstrained $(2.742)$ & unconstrained $(1.297)$ & unconstrained $(2.8)$ \\
    \hline
    $H_0$ & $72.4(76.3)_{-3.1}^{+1.8}$ & $69.23(68.54)_{-1.4}^{+0.62}$ & $68.8(70.6)_{-1.1}^{+0.5}$ \\
    $100\omega_b$ & $2.191(2.235)_{-0.038}^{+0.047}$ & $2.247(2.259)\pm 0.033$ & $2.258(2.266)_{-0.020}^{+0.018}$ \\
    $\omega_{\mathrm{cdm}}$ & $0.1356(0.1518)_{-0.013}^{+0.0062}$ & $0.1219(0.1186)_{-0.0053}^{+0.0015}$ & $0.1227(0.1281)_{-0.0036}^{+0.0018}$ \\
    $10^9A_s$ & $2.177(2.237)_{-0.079}^{+0.070}$ & $2.115(2.135)\pm0.060$ & $2.122(2.135)\pm 0.032$ \\
    $n_s$ & $0.983(1.006)_{-0.020}^{+0.015}$ & $0.9724(0.9701)_{-0.0084}^{+0.0063}$ & $0.9734(0.9823)_{-0.0076}^{+0.0053}$ \\
    $\tau_{\mathrm{reio}}$ & $0.059(0.059)\pm0.014$ & $0.057(0.065)\pm0.014$ & $0.0570(0.0574)_{-0.0076}^{+0.0069}$ \\
    \hline
    $S_8$ & $0.854(0.890)\pm 0.035$ & $0.816(0.811)_{-0.023}^{+0.018}$ & $0.831(0.839)_{-0.013}^{+0.011}$ \\
    $\Omega_m$ & $0.3015(0.3005)\pm0.0075$ & $0.3026(0.3018)\pm0.0073$ & $0.3084(0.3041)\pm0.0058$ \\
    \hline
    \end{tabular}
    \caption{The mean (best-fit) $\pm1\sigma$ errors of cosmological parameters in axion-like EDE with respect to different datasets. For upper limits, we quote the one-sided $95\%$ confidence level ($2\sigma$). The result of Planck+BAO+Pantheon is from Ref.~\cite{Simon:2022adh}.\label{tab:axion}}
\end{table}

\begin{figure}
    \centering
    \includegraphics[width=\linewidth]{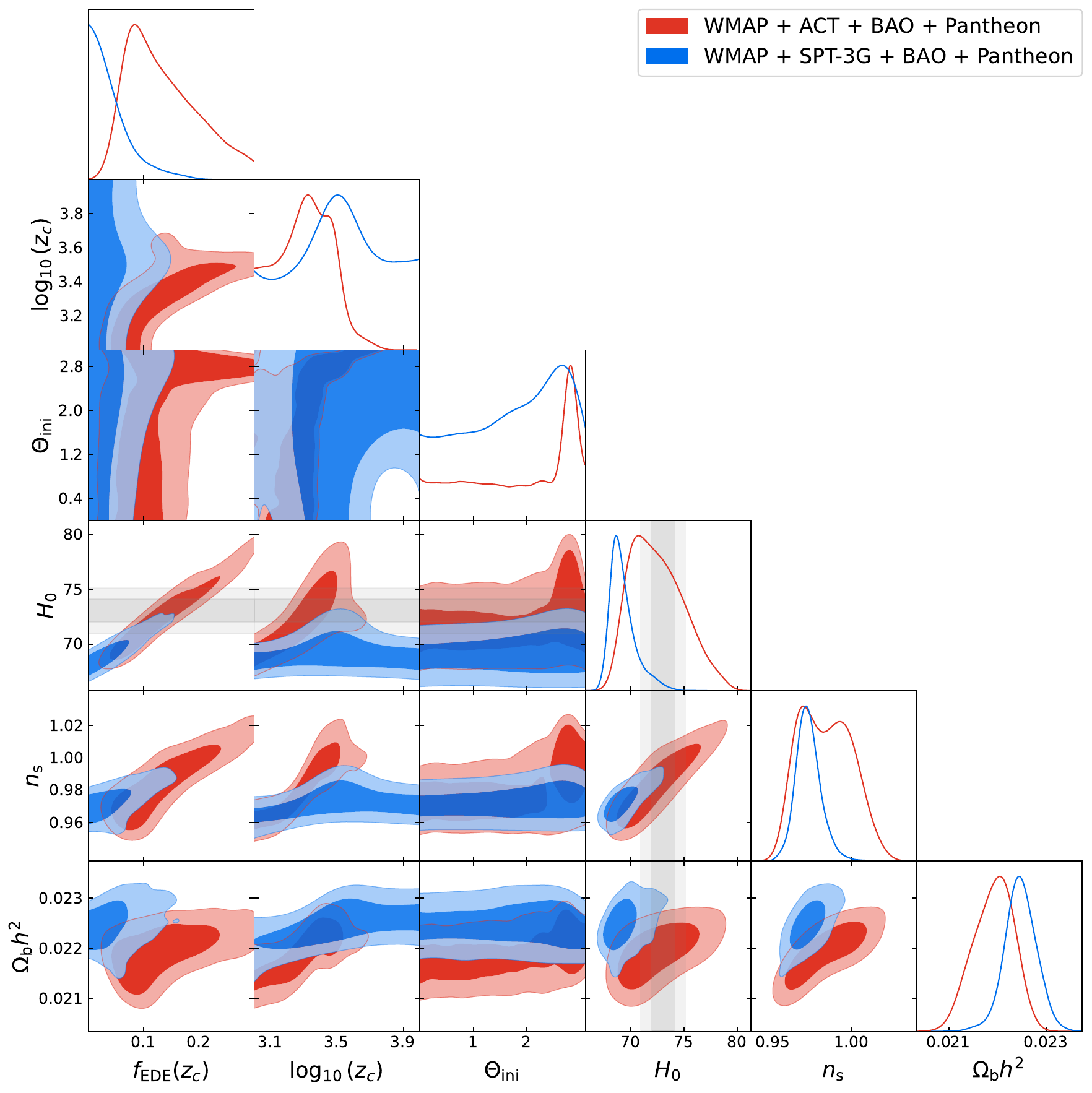}
    \caption{Marginalized posterior distributions ($68\%$ and $95\%$ confidence range) of relevant parameters in axion-like EDE fit to WMAP+ACT+BAO+Pantheon and WMAP+SPT-3G+BAO+Pantheon, respectively. The gray bands represent $1\sigma$ and $2\sigma$ regions of the latest SH0ES result\cite{Riess:2021jrx}.\label{fig:axion}}
\end{figure}

The mean (best-fit) $\pm1\sigma$ errors of cosmological parameters with respect to WMAP+ACT+BAO+Pantheon and WMAP+SPT-3G+BAO+Pantheon datasets, respectively, are showed in Table \ref{tab:axion}. We also present the result of Planck+BAO+Pantheon from Ref.~\cite{Simon:2022adh} as a comparison. The posterior distributions of relevant parameters are plotted in Fig.~\ref{fig:axion}.

The WMAP+ACT dataset prefers a non-zero amount of EDE, $f_{\mathrm{EDE}}=0.133_{-0.082}^{+0.039}$, and a large Hubble constant, $H_0=72.4_{-3.1}^{+1.8}\ \unit{km/s/Mpc}$, which is compatible with the local $H_0$ measurement. This is consistent with that in Ref.~\cite{Poulin:2021bjr}\footnote{Our result is slightly different from the result in Ref.~\cite{Poulin:2021bjr}, using the same dataset. This difference is mainly due to the different priors we choose for parameters $\log_{10}(z_c)$ and $\tau_{\mathrm{reio}}$.}, and PlanckTT($\ell_{\mathrm{max}}=650$)+ACT in Ref.~\cite{Hill:2021yec}. However, when considering the WMAP+SPT-3G dataset, we only have a upper limit on the fraction of EDE, $f_{\mathrm{EDE}}<0.119$ with the Hubble constant $H_0=69.23_{-1.4}^{+0.62}\ \unit{km/s/Mpc}$, which is consistent with Planck.

It is noteworthy that the WMAP+ACT dataset favors a lower critical redshift, $\log_{10}(z_c)=3.30_{-0.13}^{+0.20}$, so lower values of $n_s$ and $\omega_b$. In contrast, the WMAP+SPT-3G shows results consistent with Planck. We will discuss this contrast further in section-\ref{sec:ns}.

\subsection{AdS-EDE}
\begin{table}
    \centering
    \begin{tabular}{|c|c|c|c|}
    \hline
    Parameters & \makecell[c]{WMAP+ACT\\+BAO+Pantheon} & \makecell[c]{WMAP+SPT-3G\\+BAO+Pantheon} & \makecell[c]{Planck\\+BAO+Pantheon} \\
    \hline\hline
    $f_{\mathrm{EDE}}$ & $0.138(0.110)_{-0.041}^{+0.022}$ & $0.0766(0.0791)_{-0.032}^{+0.0096}$ & $0.1124(0.1084)_{-0.0070}^{+0.0038}$ \\
    $\log_{10}(z_c)$ & $3.271(3.259)_{-0.024}^{+0.086}$ & unconstrained ($3.484$) & $3.541(3.538)_{-0.036}^{+0.033}$ \\
    \hline
    $H_0$ & $73.0(72.04)_{-1.9}^{+1.5}$ & $70.14(70.77)_{-1.7}^{+0.87}$ & $72.52(72.46)\pm 0.51$ \\
    $100\omega_b$ & $2.190(2.173)_{-0.037}^{+0.058}$ & $2.290(2.293)\pm0.028$ & $2.341(2.331)_{-0.016}^{+0.018}$ \\
    $\omega_{\mathrm{cdm}}$ & $0.1379(0.1326)_{-0.0078}^{+0.0058}$ & $0.1249(0.1263)_{-0.0064}^{+0.0029}$ & $0.1346(0.1336)_{-0.0018}^{+0.0016}$ \\
    $10^9A_s$ & $2.174(2.237)\pm0.067$ & $2.128(2.134)\pm0.064$ & $2.175(2.159)\pm0.033$ \\
    $n_s$ & $0.980(0.971)_{-0.0096}^{+0.018}$ & $0.9773(0.9832)_{-0.0092}^{+0.0076}$ & $0.9964(0.9949)_{-0.0041}^{+0.0047}$ \\
    $\tau_{\mathrm{reio}}$ & $0.055(0.052)\pm0.014$ & $0.057(0.055)\pm0.015$ & $0.0545(0.0523)_{-0.0079}^{+0.0071}$ \\
    \hline
    $S_8$ & $0.865(0.842)_{-0.028}^{+0.034}$ & $0.824(0.828)_{-0.027}^{+0.022}$ & $0.863(0.856)\pm0.011$ \\
    $\Omega_m$ & $0.3008(0.2986)\pm0.0074$ & $0.3018(0.2997)\pm0.0076$ & $0.3016(0.3002)\pm0.0051$ \\
    \hline
    \end{tabular}
    \caption{The mean (best-fit) $\pm1\sigma$ errors of cosmological parameters in AdS-EDE with respect to different datasets. The result of Planck+BAO+Pantheon is from Ref.~\cite{Jiang:2021bab}.\label{tab:ads}}
\end{table}

\begin{figure}
    \centering
    \includegraphics[width=0.9\linewidth]{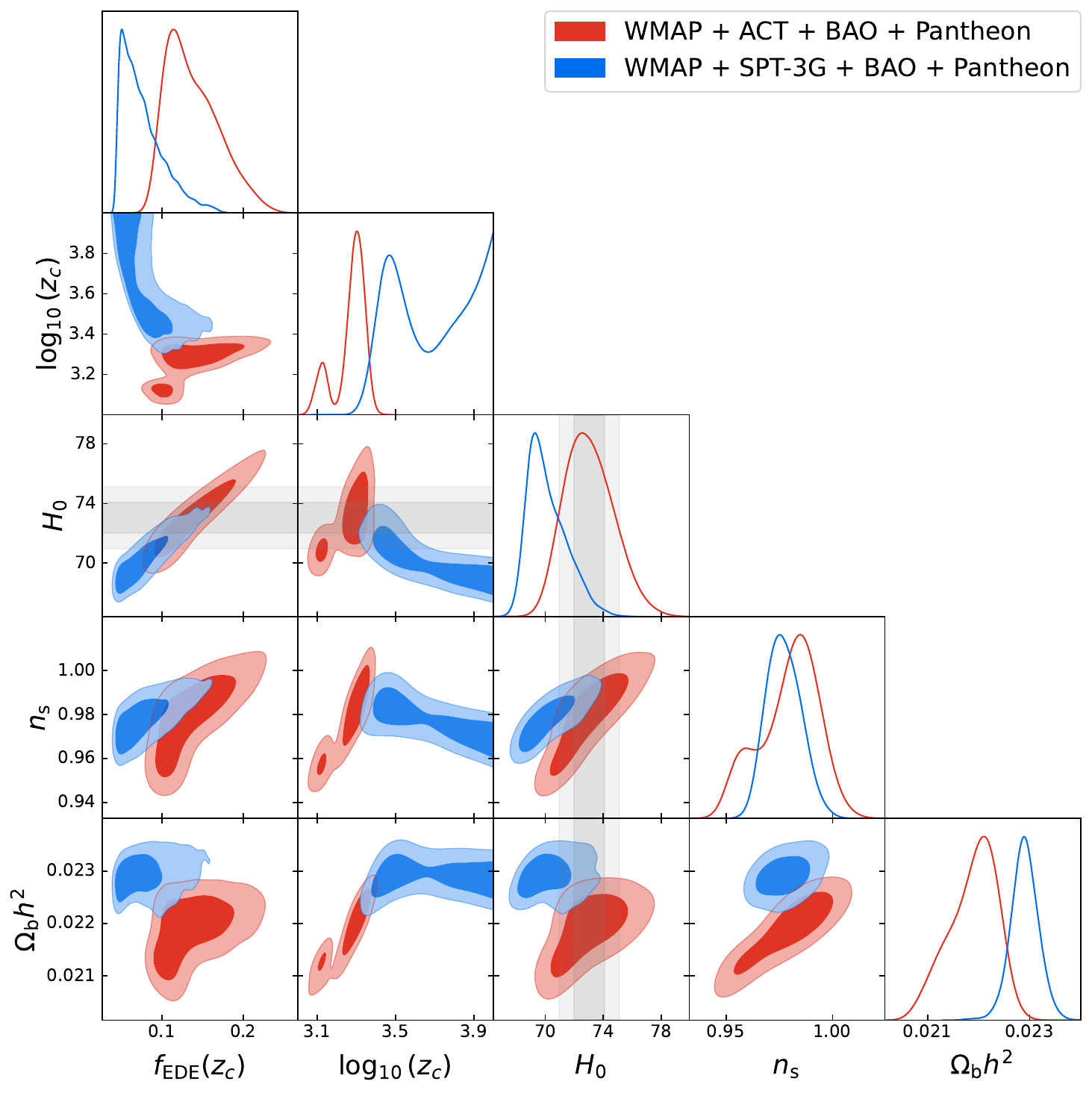}
    \caption{Marginalized posterior distributions ($68\%$ and $95\%$ confidence range) of relevant parameters in AdS-EDE fit to WMAP+ACT+BAO+Pantheon and WMAP+SPT-3G+BAO+Pantheon, respectively. The gray bands represent $1\sigma$ and $2\sigma$ regions of the latest SH0ES result\cite{Riess:2021jrx}.\label{fig:ads}}
\end{figure}

The mean (best-fit) $\pm1\sigma$ errors of cosmological parameters with respect to WMAP+ACT+BAO+Pantheon and WMAP+SPT-3G+BAO+Pantheon datasets, respectively, are showed in Table \ref{tab:ads}. We also present the result of Planck+BAO+Pantheon from Ref.~\cite{Jiang:2022uyg} as a comparison. The posterior distributions of relevant parameters are plotted in Fig.~\ref{fig:ads}.

It is well-known that AdS-EDE can have a large Hubble constant when confronted with Planck+BAO+Pantheon, without any $H_0$ prior. The result is similar (albeit with larger error bars), when we consider the WMAP+ACT dataset, which shows $f_{\mathrm{EDE}}=0.138_{-0.041}^{+0.022}$ and a Hubble constant compatible with the SH0ES result, $H_0=73.0_{-1.9}^{+1.5}\ \unit{km/s/Mpc}$. Again, as in axion-like EDE, we also find that WMAP+ACT shows smaller values for $\log_{10}(z_c)$, $n_s$ and $\omega_b$, compared to Planck.

However, the WMAP+SPT-3G dataset shows $f_{\mathrm{EDE}}=0.0766_{-0.032}^{+0.0096}$ and $H_0=70.14_{-1.7}^{+0.87}\ \unit{km/s/Mpc}$, which is worse than the results of Planck or WMAP+ACT and is in slight tension of $2.1\sigma$ with the SH0ES result\footnote{Here, $f_{\mathrm{EDE}}$ from WMAP+SPT-3G, although relatively small, is still larger than that obtained in axion-like EDE. This is probably due to the fact that we have fixed $\alpha_{\mathrm{AdS}}$, which actually sets a lower bound for $f_{\mathrm{EDE}}$ (otherwise the EDE field would not be able to climb out of the AdS well). Nevertheless, we contend that this issue is not so worrying, since the best-fit value of $f_{\mathrm{EDE}}$ is not close to the bound.}. This result is related with a higher value of critical redshift $z_c$ around $z_c\sim 10^4$, see Fig.~\ref{fig:ads}. We will discuss it further in section-\ref{sec:less_prefer}.

\section{Discussion\label{sec:discussion}}

\subsection{Difference in the fit of EDE models between datasets}
\begin{table}
    \centering
    \begin{tabular}{|c|c|c|c|}
    \hline
    Data set & $\Lambda$CDM & axion-like EDE & AdS-EDE \\ \hline
    WMAP DR5 & 5547.12 & 5544.57 & 5549.17 \\
    ACT DR4 & 292.17 & 278.46 & 276.08 \\
    BAO low-$z$ & 1.61 & 2.04 & 2.39 \\
    BAO BOSS DR12 & 3.57 & 3.40 & 3.52 \\
    Pantheon & 1034.86 & 1034.74 & 1034.73 \\ \hline
    Total $\chi^2$ & 6879.33 & 6863.20 & 6865.89 \\
    $\Delta\chi^2$ & 0 & $-16.13$ & $-13.44$ \\ \hline
    \end{tabular}
    \caption{$\chi^2$ values for the best-fit models when fit to WMAP+ACT+BAO+Pantheon dataset.\label{tab:ACT}}
\end{table}

\begin{table}
    \centering
    \begin{tabular}{|c|c|c|c|}
    \hline
    Data set & $\Lambda$CDM & axion-like EDE & AdS-EDE \\ \hline
    WMAP DR5 & 5543.27 & 5543.59 & 5542.75 \\
    SPT-3G Y1 & 1877.00 & 1876.26 & 1876.49 \\
    BAO low-$z$ & 1.59 & 1.83 & 2.02 \\
    BAO BOSS DR12 & 3.72 & 3.52 & 3.48 \\
    Pantheon & 1034.83 & 1034.75 & 1034.73 \\ \hline
    Total $\chi^2$ & 8460.40 & 8459.95 & 8459.47 \\
    $\Delta\chi^2$ & 0 & $-0.45$ & $-0.93$ \\ \hline
    \end{tabular}
    \caption{$\chi^2$ values for the best-fit models when fit to WMAP+SPT-3G+BAO+Pantheon dataset.\label{tab:SPT}}
\end{table}

In this subsection, before investigating the $n_s-H_0$ scaling relation, we check whether the axion-like EDE and AdS-EDE models show better fits to our datasets than $\Lambda$CDM. We present the best-fit $\chi^2$ per experiment when confronted with WMAP+ACT+BAO+Pantheon and WMAP+SPT-3G+BAO+Pantheon, respectively, in Table \ref{tab:ACT} and Table \ref{tab:SPT}.

It can be seen in Table \ref{tab:ACT} that both axion-like EDE and AdS-EDE have a significantly better fit than $\Lambda$CDM when fit to the WMAP+ACT dataset, and such improvements are mainly driven by ACT. In contrast, when fit to the WMAP+SPT-3G dataset, both EDE models have similar $\chi^2$ as $\Lambda$CDM but at the cost of adding extra parameters, see Table \ref{tab:SPT}.

\begin{figure}
    \centering
    \includegraphics[width=\linewidth]{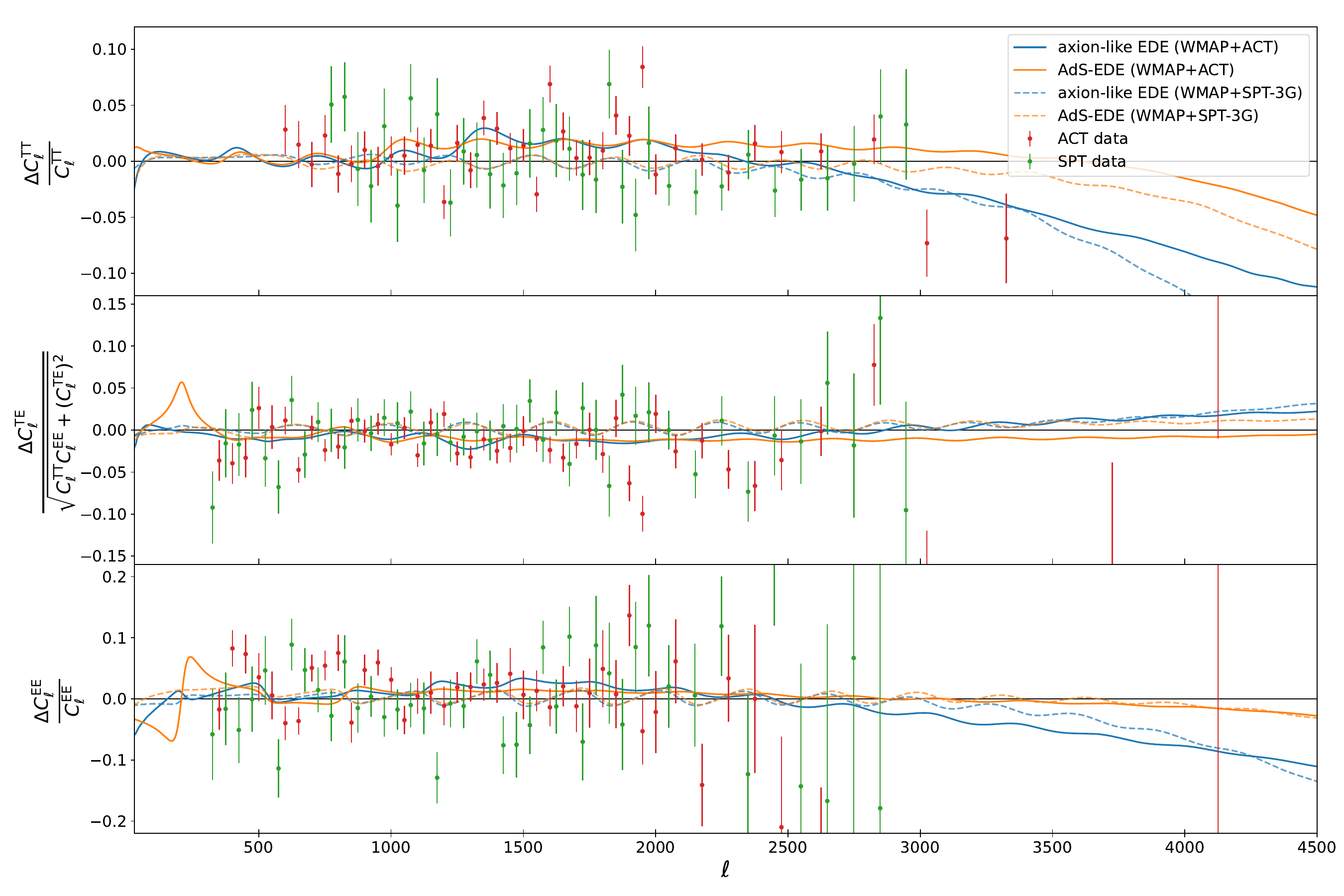}
    \caption{The relative residuals of the axion-like EDE and AdS-EDE models when fit to the WMAP+ACT dataset (solid lines) and the WMAP+SPT-3G dataset (dashed lines), respectively. The reference line is the Planck best-fit $\Lambda$CDM model from Ref.~\cite{Planck:2018vyg}. We also show the relative residuals of the ACT and SPT-3G data points with respect to the Planck best-fit $\Lambda$CDM model.\label{fig:residuals}}
\end{figure}

To see more clearly, we display in Fig.~\ref{fig:residuals} the relative residuals of both EDE models when fit to different datasets, respectively, where the reference line is the Planck best-fit $\Lambda$CDM model \cite{Planck:2018vyg}. In Fig.~\ref{fig:residuals}, we see that with WMAP+ACT, the best-fit axion-like EDE and AdS-EDE (solid lines) are favored, while with WMAP+SPT-3G, both EDE models (dashed lines) are not significantly favored.

\begin{figure}
    \centering
    \includegraphics[width=\linewidth]{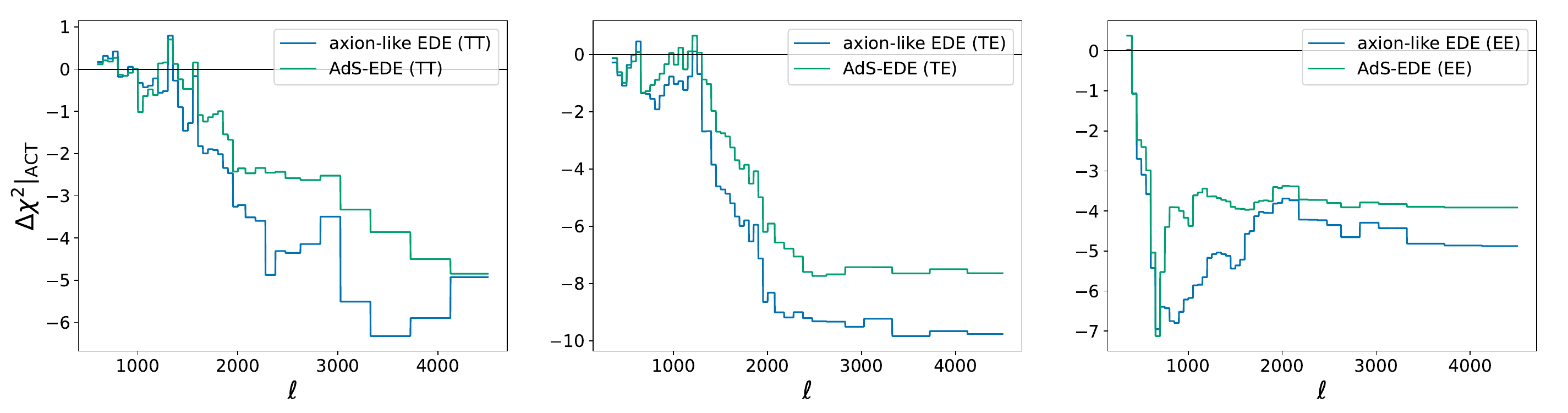}
    \caption{The cumulative $\Delta\chi^2$ of ACT for the best-fit EDE models relative to the best-fit $\Lambda$CDM model, when fit to WMAP+ACT+BAO+Pantheon. The left panel shows the TT spectrum, the middle panel shows the TE spectrum, while the right panel shows the EE spectrum.\label{fig:cumulative} }
\end{figure}

We also plot the cumulative $\Delta\chi^2$ of ACT for both EDE models in Fig.~\ref{fig:cumulative}. As observed, the improvement in $\Delta\chi^2$ for both EDE models comes from the joint contribution of the TT, TE, and EE power spectra of ACT. 
Specifically, it is found that EDE models exhibit higher values in the middle range ($1000\lesssim\ell\lesssim 2000$) of the TT spectrum and lower values at the large-$\ell$ tail ($\ell\gtrsim 3000$), which are roughly captured by the ACT data\footnote{The values of the rightmost few ACT TT bins are so low that they cannot be shown in the plot.}. 
It is worth noting that the EDE models fit to WMAP+ACT in Ref.~\cite{Poulin:2021bjr} also exhibit similar high values in the middle range of the TT power spectrum.
In addition, as also mentioned in Ref.~\cite{Hill:2021yec}, the ACT data match the lower power of the EDE models around $1200\lesssim\ell\lesssim 2000$ in the TE spectrum as well as match the oscillatary behaviour of EDE models around $l\sim 500$ in EE spectrum.

It is also noted that the axion-like EDE and AdS-EDE bring very different large-$\ell$ tails in the TT and EE spectra, but the current ACT data are not precise enough on these scales ($\ell\gtrsim 3000$) to distinguish between different EDE models.

\subsection{Towards an understanding of $n_s-H_0$ scaling relation \label{sec:ns}}

\begin{figure}
    \centering
    \includegraphics[width=\linewidth]{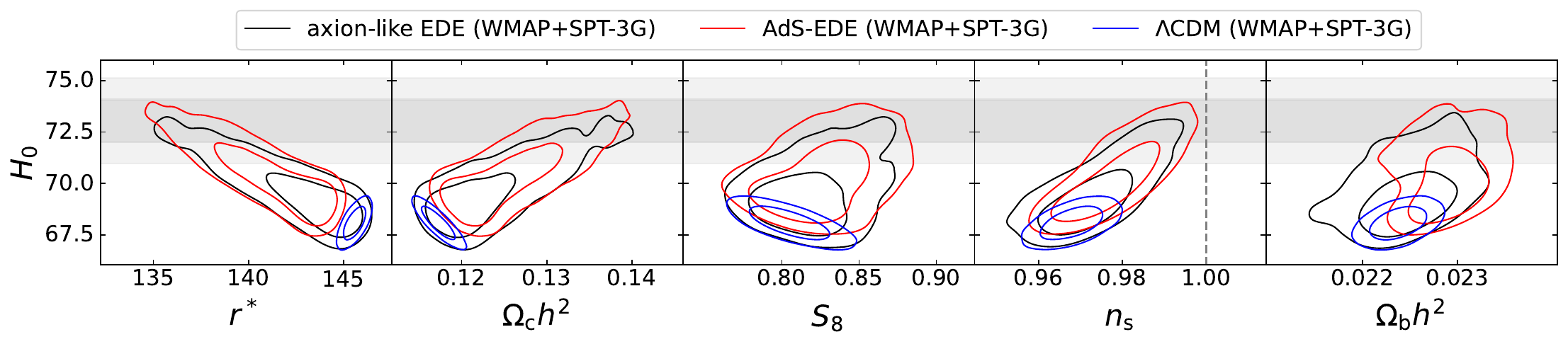}
    \caption{The degeneracy of paramteters for models fit to WMAP+SPT-3G+BAO+Pantheon dataset.\label{fig:parameters1}}
\end{figure}

\begin{figure}
    \centering
    \includegraphics[width=\linewidth]{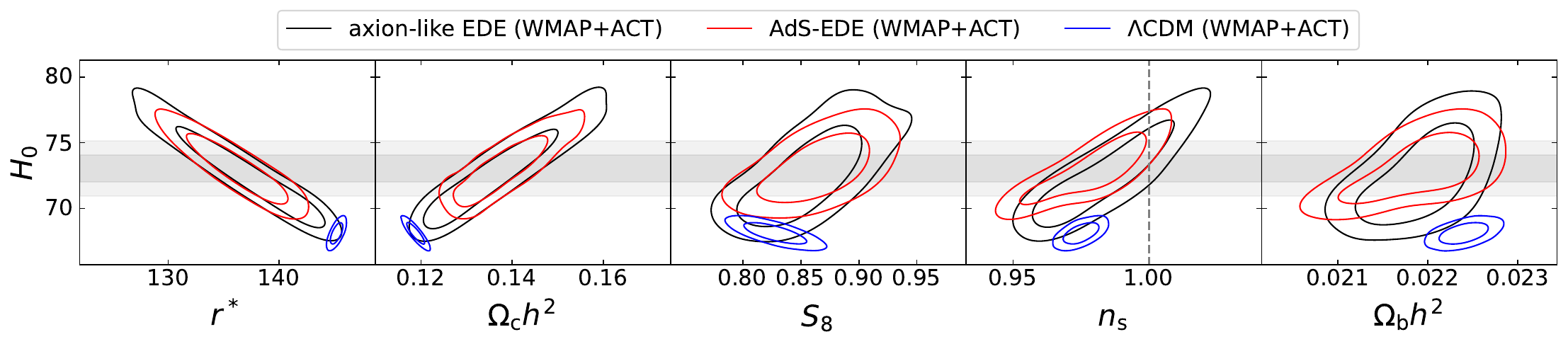}
    \caption{The degeneracy of paramteters for models fit to WMAP+ACT+BAO+Pantheon dataset.\label{fig:parameters2}}
\end{figure}

It has been elaborated in Ref.~\cite{Ye:2021nej} that any pre-combination resolution of the Hubble tension, including EDE, inevitably causes the shifts of other cosmological parameters. As also observed in \cite{Vagnozzi:2021gjh}, a higher value of $\omega_{\mathrm{cdm}}$ is required to offset the increment in the early integrated Sachs–Wolfe (ISW) effect due to the existence of EDE, which thus leads to a higher $S_8$. 
The shifts of relevant parameters can be seen in our results presented in Fig.~\ref{fig:parameters1} and Fig.~\ref{fig:parameters2}.

In particular, Ref.~\cite{Ye:2021nej} pointed out that the spectrum index $n_s$ and baryon density $\omega_b$ increase linearly with $H_0$ in any pre-combination resolution of the Hubble tension. Since the angular scale of damping 
\begin{equation} \label{eq:damping}
    \theta_D^*=\frac{r_D^*}{D_A^*} \sim r_D^*H_0 \sim \omega_b^{-1/2}\omega_{\mathrm{cdm}}^{-1/4}H_0 
\end{equation}  
is constrained by the CMB data, and $\omega_{\mathrm{cdm}}H_0^{-2}$ is fixed by the CMB and BAO data, we have $\omega_b^{-1}H_0\sim\mathrm{const}$. Therefore, a larger $\omega_b$ is needed to offset the effect of higher $H_0$, and also a higher $n_s$ is required to compensate for the enhanced baryon loading effect due to a higher $\omega_b$. Consequently, Ref.~\cite{Ye:2021nej} suggested an universal $n_s-H_0$ scaling relation:
\begin{equation}
    \delta n_s \simeq 0.8(1-\alpha) \frac{\delta H_0}{H_0}
\end{equation}
where $\alpha$ is a parameter that marginalizes the unclear extra damping needed to compensate for a larger $n_s$.

\begin{figure}
    \centering
    \includegraphics[width=0.4\linewidth]{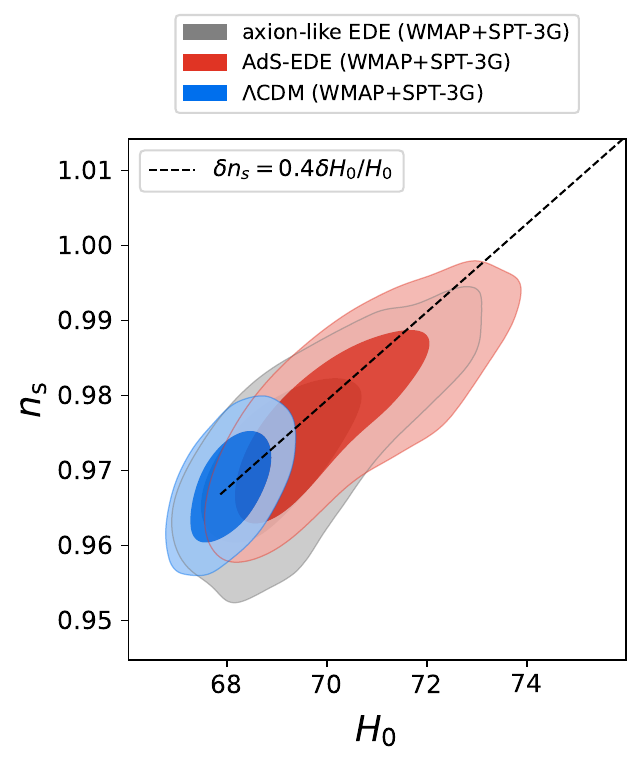}
    \qquad
    \includegraphics[width=0.4\linewidth]{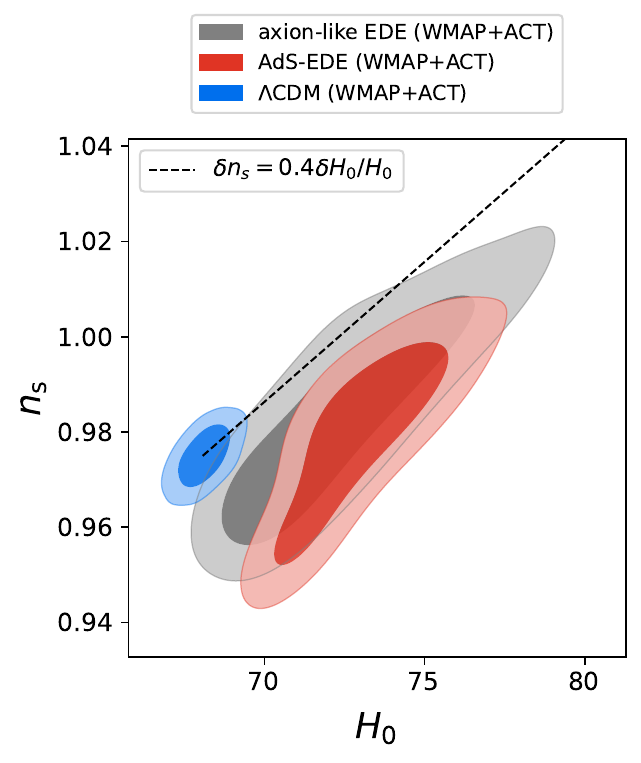}
    \caption{The $n_s-H_0$ scaling relation for different models fit to  datasets WMAP+SPT-3G (left) and WMAP+ACT (right), respectively. The dashed lines represent ${\delta n_s}= 0.4\frac{\delta H_0}{H_0}$ and start from the best-fit values of $\Lambda$CDM fit to corresponding datasets. \label{fig:nsH0}}
\end{figure}

Specifically, Ref.~\cite{Ye:2021nej} further found that for the Planck+BAO+Pantheon+R19 dataset, the spectral index will scale linearly with the Hubble constant as (\ref{eq:deltans}) ($\alpha\simeq 0.5$).
As a result, a Hubble constant around $H_0\simeq \qty{73}{km/s/Mpc}$ would correspond to a Harrison-Zeldovich spectrum ($n_s=1$). It is interesting to see in the left panel of Fig.~\ref{fig:nsH0} that the WMAP+SPT-3G dataset, despite favoring lower $H_0$ due to the absence of low-redshift $H_0$ prior, also shows such a linear $n_s$-$H_0$ scaling relation of (\ref{eq:deltans}).

However, as can be seen from the right panel of Fig.~\ref{fig:nsH0}, the WMAP+ACT dataset prefers lower $\omega_b$ and $n_s$ but higher $H_0$ for axion-like EDE and AdS-EDE, making the $n_s$-$H_0$ relation seem to deviate from linearity and inconsistent with (\ref{eq:deltans}).
In particular, with WMAP+ACT+BAO+Pantheon, we can only obtain $n_s\simeq 0.98$ for $H_0\simeq \qty{73}{km/s/Mpc}$, while we have $n_s\simeq 1$ for $H_0\simeq \qty{73}{km/s/Mpc}$ with Planck+BAO+Pantheon or Planck($\ell_{TT}\lesssim 1000$)+ACT+SPT+BAO+Pantheon \cite{Jiang:2022uyg}.

It should be noted that the arguments in Ref.~\cite{Ye:2021nej} are based on the assumption that the exciting of EDE is near the matter-radiation equality, i.e.$z_c\sim z_{\mathrm{eq}}$, and thus has a negligible effect on the damping scale $r_D^*$. It is under this premise that Eq.~(\ref{eq:damping}) is a good approximation, leading to the result that $\omega_b$ and $n_s$ increase linearly with $H_0$. However, as mentioned above, our results with WMAP+ACT prefer a lower $z_c$ (closer to the recombination time), such that the injection of EDE will non-negligibly suppress the damping scale $r_D^*$, and so $\omega_b$ and $n_s$. It is worth noting that similar results for $\omega_b$ and $n_s$ in EDE models can also be seen in Ref.~\cite{Hill:2021yec} with PlanckTT($\ell_{\mathrm{max}}=650$)+ACT dataset.

\begin{figure}
    \centering
    \includegraphics[width=0.9\linewidth]{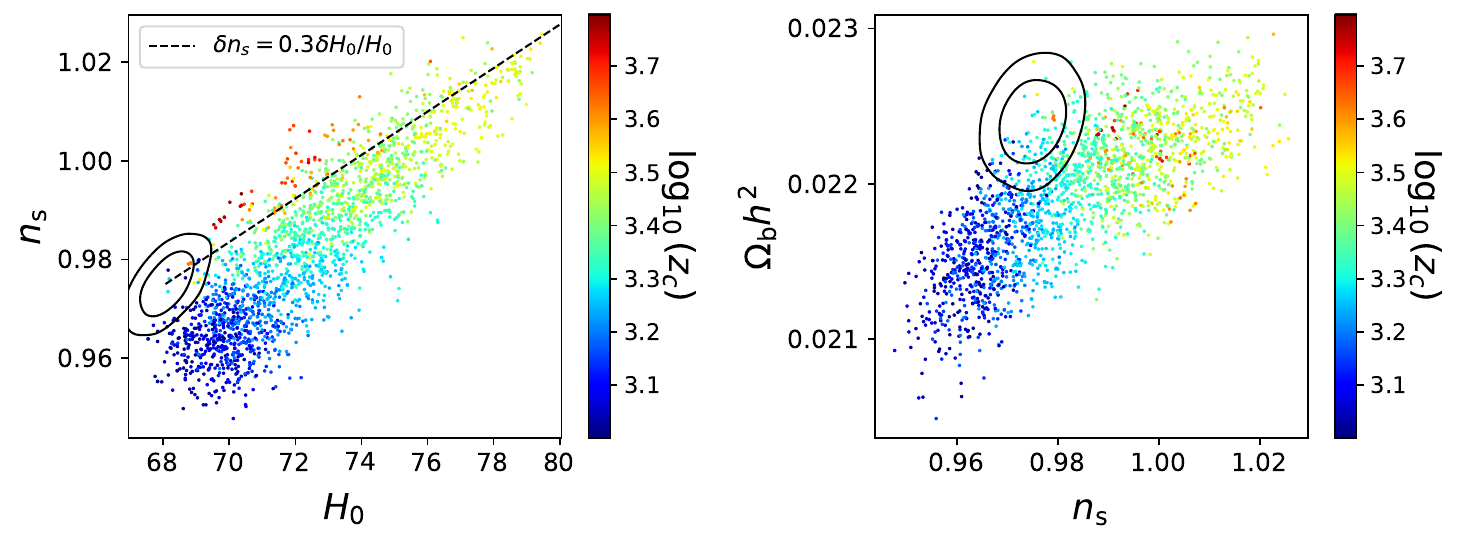}
    \caption{Scatter plots of relevant parameters in axion-like EDE fit to WMAP+ACT+BAO+Pantheon dataset. The left panel is $H_0$-$n_s$ plot, while the right panel is $n_s$-$\omega_b$ plot, both with color coding for $\log_{10}(z_c)$. We also show the contours of $\Lambda$CDM as comparisons. The dashed line in the left panel corresponds to ${\delta n_s}= 0.3\frac{\delta H_0}{H_0}$. The result is similar for AdS-EDE.\label{fig:ns}}
\end{figure}

To see this more clearly, we present scatter plots of relevant parameters with color coding for $\log_{10}(z_c)$ in Fig.~\ref{fig:ns}, which show that lower values of $n_s$ and $\omega_b$ indeed correspond to lower $z_c$. It can be seen in the left panel of Fig.~\ref{fig:ns} that if we only consider the points around the matter-radiation equality time, i.e. $\log_{10}(z_c)\sim 3.5$, we can again have a scaling relation similar to (\ref{eq:deltans}), except with a slightly smaller scale factor:
\begin{equation}
    {\delta n_s}\simeq 0.3\frac{\delta H_0}{H_0},
\end{equation}
which is consistent with the scaling relation obtained in Ref.~\cite{Jiang:2022uyg} using Planck($\ell_{TT}\lesssim 1000$)+ACT+SPT+BAO+Pantheon. However, if we consider the points around $\log_{10}(z_c)\sim 3.3$, the best-fit $z_c$ to WMAP+ACT, smaller values of $n_s$ are favored. Therefore, we can conclude that, when fit to WMAP+ACT, the smaller $n_s$ in EDE models is correlated with the lower critical redshift $z_c$.

\subsection{The less preference of WMAP+SPT-3G for AdS-EDE\label{sec:less_prefer}}

AdS-EDE has the advantage of yielding a large Hubble constant, $H_0\sim \qty{73}{km/s/Mpc}$, when fit to Planck, even without the inclusion of any $H_0$ prior \cite{Ye:2020btb,Ye:2020oix,Jiang:2021bab}. This result also holds for WMAP+ACT. However, we have found that WMAP+SPT-3G seems to have relatively less support for AdS-EDE, with a lower $H_0=70.14_{-1.7}^{+0.87}\ \unit{km/s/Mpc}$, which might be partly due to higher $z_c$ values (preferred by WMAP+SPT-3G) since a high critical redshift around $z_c\sim 10^4$ is unphysical in resolving the Hubble tension\footnote{In order to resolve the Hubble tension, we require EDE to be relevant during a short epoch around the matter-radiation equality ($\log_{10}(z_c)\sim 3.5$) and thus reduce the sound horizon $r_s^*$. However, if the critical redshift $z_c$ is too high, which means EDE starts to decay too early, EDE becomes negligible before significantly reducing the sound horizon $r_s^*$.}. 
We can see this in the $f_{\mathrm{EDE}}$-$\log_{10}(z_c)$ plot of Fig.~\ref{fig:ads} that a higher $z_c$ corresponds to a lower $f_{\mathrm{EDE}}$, so a lower $H_0$.

\begin{figure}
    \centering
    \includegraphics[width=0.6\linewidth]{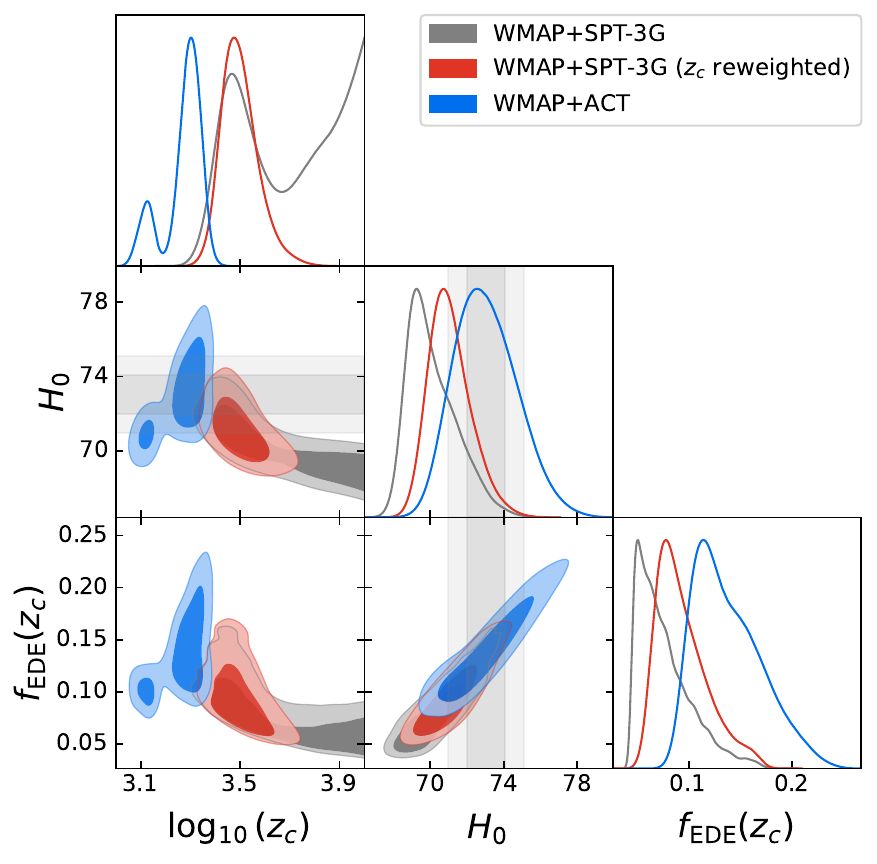}
    \caption{Marginalized posterior distributions ($68\%$ and $95\%$ confidence range) of relevant parameters in AdS-EDE. The red contours represent the result of WMAP+SPT-3G with a Gaussian weight on critical redshift: $\log_{10}(z_c)=3.5\pm0.1$, while the grey contours represent the original WMAP+SPT-3G result. We also plot the result of WMAP+ACT (blue) as a comparison.\label{fig:reweight}}
\end{figure}

To study the impact of high $z_c$ values, we follow Ref.~\cite{LaPosta:2021pgm} and perform a re-weighting of the samples using a Gaussian weight on critical redshift: $\log_{10}(z_c)=3.5\pm0.1$, with which we can focus only on the samples around the matter-radiation equality and exclude those with higher $z_c$ values. The result is shown in Fig.~\ref{fig:reweight}. It is found that $f_{\mathrm{EDE}}=0.094_{-0.030}^{+0.014}$ and $H_0=71.07_{-1.3}^{+0.89}\ \unit{km/s/Mpc}$ for the reweighted samples, which is slightly larger than the original WMAP+SPT-3G result, but still inconsistent with SH0ES by $1.4\sigma$.

\section{conclusion\label{sec:conclusions}}

In this work, based on axion-like EDE and AdS-EDE, we have investigated the $n_s-H_0$ scale relation with Planck-independent CMB data, i.e. ACT and SPT-3G combined respectively with WMAP. The main results are as follows:
\begin{itemize}
    \item The WMAP+ACT dataset favors a non-zero amount of EDE, without any late-time $H_0$ prior, which is consistent with Refs.\cite{Hill:2021yec,Poulin:2021bjr} that considered ACT DR4. In contrast, the WMAP+SPT-3G dataset cannot give a stronger preference for the EDE models than Planck, which leads only to an upper limit on $f_{\mathrm{EDE}}$ for axion-like EDE, and gives $f_{\mathrm{EDE}}=0.0766_{-0.032}^{+0.0096}$ and $H_0=70.14_{-1.7}^{+0.87}\ \unit{km/s/Mpc}$ for AdS-EDE. 
    \item The WMAP+SPT-3G dataset also follows the $n_s$-$H_0$ scaling relation shown in Eq.~(\ref{eq:deltans}). However, the WMAP+ACT dataset prefers smaller values of $n_s$, which is related to the fact that the critical redshift $z_c$ favored by this dataset is lower and closer to the recombination time, while (\ref{eq:deltans}) requires that the exciting of EDE is near the matter-radiation equality, i.e. $z_c\simeq z_{\mathrm{eq}}$.
\end{itemize}

The $n_s-H_0$ scaling relation (\ref{eq:deltans}) has significant implications for our insight into inflation and the primordial Universe, see e.g. \cite{Kallosh:2022ggf,Ye:2022efx,Takahashi:2021bti,DAmico:2021fhz,Braglia:2022phb,Giare:2023kiv,Huang:2023chx}. Based on the above results and previous studies\cite{Hill:2021yec,Poulin:2021bjr}, we can observe that the ACT-DR4 data gives slightly different results from Planck in both the preference of EDE and the $n_s-H_0$ relation. However, it has been pointed out that the ACT DR4 data has some mild discrepancies with other CMB datasets \cite{ACT:2020gnv,Handley:2020hdp,DiValentino:2022rdg,Giare:2022rvg,Calderon:2023obf}. Thus with the arrival of the next-generation CMB surveys, such as Simons Observatory\cite{SimonsObservatory:2018koc} and CMB-S4\cite{Abazajian:2019eic}, it is expected that we will have a better perspective on EDE and the $n_s-H_0$ scaling relation.

It is also noted that the MCMC analysis suffers from the prior volume effects \cite{Herold:2021ksg,Gomez-Valent:2022hkb}, which can bias the posterior towards a lower EDE fraction. It has been found in Refs.~\cite{Herold:2021ksg,Herold:2022iib} that a non-zero $f_{\mathrm{EDE}}$ can be obtained with Planck+LSS data by performing a frequentist analysis based on profile likelihoods, see also Ref.~\cite{Cruz:2023cxy} for NEDE. Therefore, further testing EDE and the $n_s-H_0$ scaling relation with Planck-independent CMB data using profile likelihoods is also significant.

\begin{acknowledgments}

    We thank Jun-Qian Jiang for valuable discussion. ZYP is supported by UCAS Undergraduate Innovative Practice Project. YSP is supported by NSFC, No.12075246 and the Fundamental Research Funds for the Central Universities. We acknowledge the use of publicly available codes AxiCLASS (\url{https://github.com/PoulinV/AxiCLASS}) and classmultiscf (\url{https://github.com/genye00/class_multiscf.git}).

\end{acknowledgments}

\appendix

\section{The EDE models\label{app:models}}
In this appendix, we briefly describe the EDE models used. In EDE resolution of Hubble tension, an unknown component, i.e.EDE, behaves like a cosmological constant at $z\gtrsim 3000$ and then must decay rapidly before recombination, so that it just suppressed the sound horizon but dose not affect the late evolution of the Universe. The angular scale of sound horizon at recombination
\begin{equation}
    \theta_s^* = \frac{r_s^*}{D_A^*} \sim r_s^*H_0
\end{equation}
can be precisely set with CMB data, where $D_A^*$ is the angular diameter distance to last scattering. Therefore, if the evolution after recombination follows flat $\Lambda$CDM, thus with a lower $r_s$, we naturally obtain a higher value of $H_0$.

In this paper, we consider two well-known EDE models. The first is axion-like EDE \cite{Poulin:2018cxd,Smith:2019ihp}, which is the original model of EDE. In this model, EDE is an ultra-light scalar field $\phi$ with an axion-like potential,
\begin{equation}
V(\theta) = m^2f^2\left(1-\cos\theta\right)^n, \quad \theta \in
\left[-\pi,\pi\right]
\end{equation}
where $\theta \equiv \phi/f$ is the re-normalized field variable, $m$ and $f$ are the effective mass and the couple constant of axion-like EDE, respectively, see also \cite{McDonough:2022pku,Cicoli:2023qri} for modelling it in string theory. At early times, it is frozen at certain initial value, $\theta_i=\phi_i/f$, due to the Hubble friction, and behaves like dark energy. Afterwards, as the Hubble parameter falls, the field will start to roll down at a critical redshift $z_c$ and rapidly oscillate. As a result, the energy density of EDE will decay with an equation of state $w\approx (n-1)/(n+1)$ \cite{PhysRevD.28.1243,Poulin:2018dzj}. In this work, we will set $n=3$ following Ref.~\cite{Poulin:2018cxd}.

Another EDE model we consider is AdS-EDE \cite{Ye:2020btb}, in which we have an AdS phase around recombination. In this work, we consider a phenomenological potensial\footnote{Other potentials are also possible, see e.g. \cite{Ye:2020oix}.}
\begin{equation}
    V(\phi)= \begin{dcases}
        V_0\left(\frac{\phi}{M_{\mathrm{Pl}}}\right)^4-V_{\mathrm{AdS}}, & \frac{\phi}{M_{\mathrm{Pl}}}<\left(\frac{V_{\mathrm{AdS}}}{V_0}\right)^{1 / 4} \\ 0 , & \frac{\phi}{M_{\mathrm{Pl}}}>\left(\frac{V_{\mathrm{AdS}}}{V_0}\right)^{1 / 4}
    \end{dcases}
\end{equation}
where $V_{\mathrm{AdS}}$ is the depth of the AdS well, $M_{\mathrm{Pl}}$ is the reduced Planck mass.
The implications of AdS vacuum for our current Universe also have been studied in recent Refs.\cite{Visinelli:2019qqu,Akarsu:2019hmw,Calderon:2020hoc,Akarsu:2021fol,Akarsu:2022typ,Sen:2021wld,DiGennaro:2022ykp,Ong:2022wrs,Malekjani:2023dky,Adil:2023exv,Adil:2023ara}. The existence of an AdS phase makes the energy density of EDE decay faster than in oscillation phase. Therefore, compared to axion-like EDE, AdS-EDE can allow a more efficient injection of EDE with less influence on the fit to CMB data. As a result, AdS-EDE has the advantage of yielding a large Hubble constant, $H_0\sim \qty{73}{km/s/Mpc}$, without the inclusion of any $H_0$ prior\cite{Ye:2020btb,Ye:2020oix,Jiang:2021bab}.

\section{Results of $\Lambda$CDM}\label{app:LCDM}

\begin{table}
    \centering
    \begin{tabular}{|c|c|c|}
    \hline
    Parameters & \makecell[c]{WMAP+ACT\\+BAO+Pantheon} & \makecell[c]{WMAP+SPT-3G\\+BAO+Pantheon} \\
    \hline\hline
    $H_0$ & $68.12(68.07)\pm 0.54$ & $68.07(67.87)\pm 0.53$ \\
    $100\omega_b$ & $2.240(2.243)\pm0.018$ & $2.238(2.235)\pm0.020$  \\
    $\omega_{\mathrm{cdm}}$ & $0.1188(0.1190)\pm 0.0013$ & $0.1173(0.1178)\pm 0.0013$  \\
    $10^9A_s$ & $2.147(2.159)\pm 0.059$ & $2.097(2.107)\pm0.057$  \\
    $n_s$ & $0.975(0.975)\pm 0.0043$ & $0.9679(0.9668)\pm 0.0049$ \\
    $\tau_{\mathrm{reio}}$ & $0.063(0.066)\pm0.014$ & $0.057(0.059)\pm0.014$ \\
    \hline
    $S_8$ & $0.827(0.832)\pm 0.018$ & $0.805(0.812)\pm 0.017$ \\
    $\Omega_m$ & $0.3058(0.3066)\pm 0.0073$ & $0.3029(0.3055)\pm0.0072$ \\
    \hline
    \end{tabular}
    \caption{The mean (best-fit) $\pm1\sigma$ errors of cosmological parameters in $\Lambda$CDM with respect to different datasets.}\label{tab:lcdm}
\end{table}

We show the MCMC results of $\Lambda$CDM in Table \ref{tab:lcdm}.

\bibliography{ref}

\end{document}